\documentclass[aps,prl,twocolumn,groupedaddress]{revtex4}

\usepackage[]{graphicx}
\usepackage{amsmath}       
\usepackage{amssymb}       
\usepackage{url}

\begin{document}

\begin{widetext}
\noindent\textbf{Preprint of:}\\
Gregor Kn\"{o}ner, Simon Parkin, Timo A. Nieminen,
Norman R. Heckenberg and Halina Rubinsztein-Dunlop\\
``Measurement of refractive index of single microparticles''\\
\textit{Physical Review Letters} \textbf{97}(15), 157402 (2006)
\end{widetext}


\title{Measurement of refractive index of single microparticles}


\author{Gregor Kn\"oner}
\author{Simon Parkin}
\author{Timo A. Nieminen}
\author{Norman R. Heckenberg}
\author{Halina Rubinsztein-Dunlop}
\affiliation{Centre for Biophotonics and Laser Science, School of
Physical Sciences, The University of Queensland, Brisbane QLD 4072,
Australia}


\date{\today}

\begin{abstract}
The refractive index of single microparticles is derived from precise measurement and rigorous modeling of the stiffness of a laser trap. We demonstrate the method for particles of four different materials with diameters from 1.6 to 5.2\,$\mu$m and achieve an accuracy of better than 1\,\%. The method greatly contributes as a new characterization technique because it works best under conditions (small particle size, polydispersion) where other methods, like absorption spectroscopy, start to fail. Particles need not to be transferred to a particular fluid, which prevents particle degradation or alteration common in index matching techniques. Our results also show that advanced modeling of laser traps accurately reproduces experimental reality.
\end{abstract}


\maketitle


The refractive index of micrometer sized objects is a highly sought-after property, but it is not easily measured. It is important because it dictates particle interaction with light in systems spanning the range from materials science (e.g particles in paint, new polymers for drug delivery, phototherapy, electro-optics \cite{sumpter2003}) through atmospheric physics (light scattering by aerosol particles) to oceanography (importance of the refractive index of plankton cells \cite{stramski1999}). Recent advances in the field of biophysics, which allow the study of cell functions at a single cell level, have spawned an increased interest in characterizing and modeling the optical properties of single cells and organelles within the cell \cite{backman2000}.

Several techniques are used today for refractive index measurements. Standard refractometry is used to determine the refractive index of bulk liquid samples. Laser diffractometry is widely used to investigate light scattering by cells suspended in liquid to infer refractive index and shape \cite{bessis1980}, but is limited to large samples of monodisperse particles. Index matching is most commonly used to measure the refractive index of particles in suspension. It can not be applied to microparticles that must be maintained in a particular environment to avoid destruction, degradation, or alteration of optical properties, such as biological specimens or crystals in saturated solution.

Here we describe an accurate method for the determination of the refractive index of an individual spherical particle in an optical trap. The refractive index is determined by modeling the stiffness in the optical trap for a range of refractive indices, measuring the actual trap stiffness and comparing both results. Only a very basic single beam optical tweezers setup with a detection of forward scattered light is necessary, which allows the method to be integrated into standard research microscope as well as lab-on-a-chip applications. The method has several advantages: it measures the refractive index of one individual particle, that can simultaneously be manipulated. Particles do not have to be suspended in a special liquid. Particle sizes can be smaller than 5\,$\mu$m. A narrow size distribution of the particles in the sample is not necessary, a broad distribution is even an advantage.

These properties allow the method to be applied to a range of problems: characterization of particles in a polydisperse sample, time lapse experiments where the particle is held in the trap over longer times and the stiffness continuously measured, or application to the characterization of living cells and organelles. Immobilization of the particle of interest in the laser trap also enables the use of microfluidic devices to change the surrounding liquid or add elements like salts or biological factors to the liquid, while continually quantifying the particle's optical properties. On the other hand, the method could also be used for characterizing the optical properties of the liquid, when the particle refractive index is known.

\begin{figure}[ht!]
   \includegraphics[clip,width=0.8\hsize]{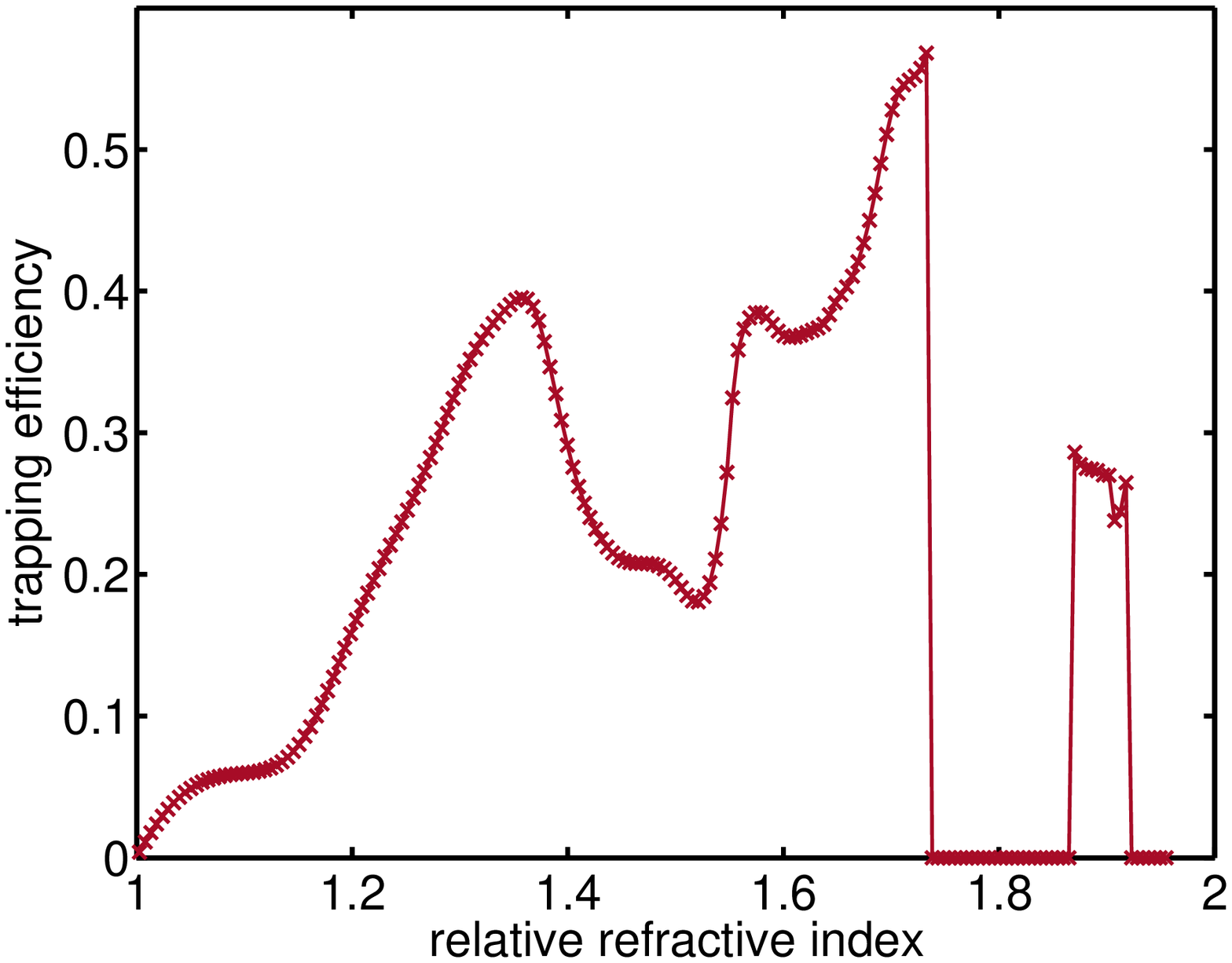}\\
   \includegraphics[clip,width=0.8\hsize]{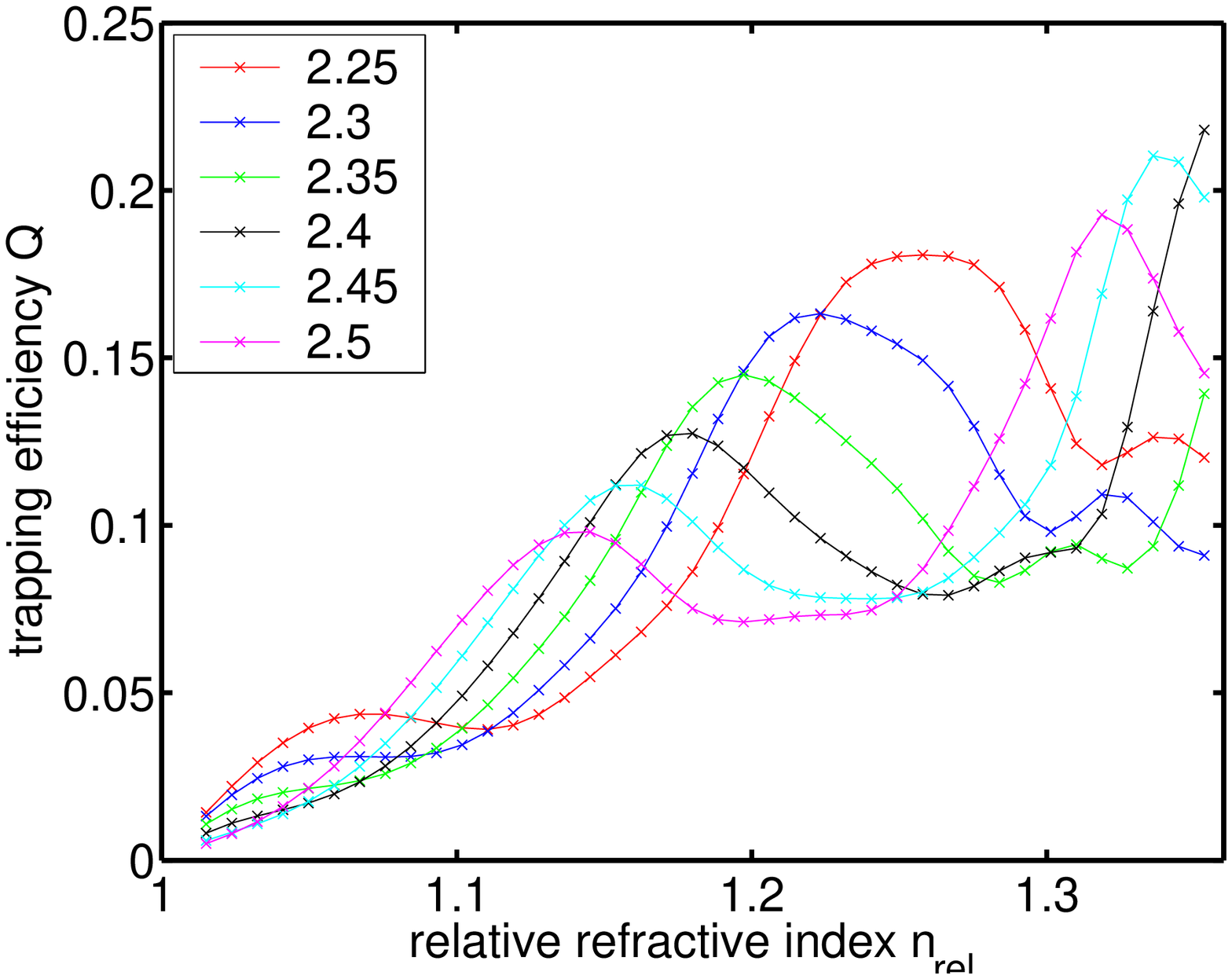}
   \caption{Upper: The trapping efficiency $Q$ of a 2.09\,$\mu$m particle increases monotonically to $n_{\mathrm{rel}}=1.36$ meaning that this is a useful range for refractive index determination.
If the scattering force, which strongly depends on the refractive index, outweighs the gradient force, stable 3D trapping is not possible. This situation is marked with values of zero efficiency.
Lower: The monotonic region shifts with particle size. Particle radii are given in terms of the wavelength $\lambda$ (3.5 to 4\,$\mu$m diameter at 1070\,nm).}
   \label{trace_index_wide}
\end{figure}

An optical trap is created by strongly focussing a laser beam with a high numerical aperture (NA) objective lens. To model the optical forces acting on a particle of a size comparable to the laser wavelength (here 1070\,nm), electromagnetic scattering theory has to be employed. The electric fields must satisfy the vector Helmholtz equation
\begin{equation}
  \nabla^2 {\bf E} + k^2 \bf{E} =  0
\end{equation}
where $k$ is the wave number.
The equations for the magnetic fields are similar and will not be mentioned explicitly here.
For a source free region, one can find a general solution for the electric field ${\bf E}$ in spherical coordinates as a superposition of the divergence free vector spherical wave functions (VSWF) ${\bf M}_{nm}$ and ${\bf N}_{nm}$:
\begin{eqnarray}
\label{field_eq}
  {\bf E} =  \sum_{n = 1}^{\infty} \sum_{m = -n}^{n} \Big( a_{nm} {\bf M}^{(2)}_{nm} & + & b_{nm} {\bf N}^{(2)}_{nm} + \nonumber\\
                                                      + p_{nm} {\bf M}^{(1)}_{nm} & + & q_{nm} {\bf N}^{(1)}_{nm} \Big)\,.
\end{eqnarray}

Incoming fields depend on the expansion coefficients $a_{nm}$ and $b_{nm}$ and the modes ${\bf M}_{nm}^{(2)}$, ${\bf N}_{nm}^{(2)}$ whereas outgoing fields depend on the coefficients $p_{nm}$ and $q_{nm}$ and the modes ${\bf M}_{nm}^{(1)}$, ${\bf N}_{nm}^{(1)}$. The VSWFs are functions of the vector spherical harmonics and the Hankel functions of the first (1) and second (2) kind and are given elsewhere \cite{nieminen2003}. The incident strongly focussed laser beam can be described as a multipole expansion with the VSWFs as a basis and the expansion coefficients $a_{nm}$ and $b_{nm}$. Experimentally, we use a collimated Gaussian beam overfilling an NA=1.3 oil immersion objective (Olympus, $100\times$, cutoff at 1/e of maximal intensity). For modeling, we project the Gaussian intensity distribution onto a spherical surface with a truncation at 1/e of the intensity at a truncation angle $\varphi$ given by NA = $n_{\mathrm{med}} \sin \varphi$ ($n_{\mathrm{med}}$ denotes the medium refractive index). The multipole expansion coefficients are then obtained by a point matching algorithm in the far field. This method is independent of the scatterer, fast and reliable \cite{nieminen2003}. Incoming and outgoing modes are related to each other by the scattering matrix or ${\bf T}$-matrix \cite{waterman1971}, which is obtained from the boundary conditions on the particle's surface
\begin{equation}
\left[ \begin{array}{c} p_{nm} \\ q_{nm} \end{array}  \right] = \bigg[ \;\; {\bf T} \;\;  \bigg] \cdot \left[ \begin{array}{c} a_{nm} \\ b_{nm} \end{array}  \right]\,.
\end{equation}
In the case of a a homogeneous isotropic spherical particle, the ${\bf T}$-matrix is diagonal. Its elements are given by the analytical Lorenz--Mie solution \cite{mie1908} and depend on the particle diameter $d$ and the relative refractive index $n_{\mathrm{rel}}=n_{\mathrm{part}}/n_{\mathrm{med}}$ as well as Hankel and Bessel functions that require numerical evaluation.

Knowledge of the total fields (Eq. \ref{field_eq}) allows calculation of the forces acting on the scatterer. We base our calculations on the conservation of momentum: The momentum transferred to the particle must equal the change of momentum in the beam. By integrating the momentum flux over a spherical surface in the far field, we obtain the change of momentum in the beam and thus the applied force ${\bf F}$. The integration can be performed analytically and reduces the calculation of the force to the summation over the expansion coefficients of the fields ($a_{nm}$, $b_{nm}$ and $p_{nm}$, $q_{nm}$) \cite{crichton2000, nieminenspie2004}.

The following procedure is used to calculate the trap stiffness. The equilibrium position of the particle in the trap is found by minimizing the axial force $F_{z}$ on the particle with a bisection search that terminates at $10^{-7}$\,pN. The transverse trap stiffness at that position is obtained by calculating the force on the particle for a small transverse displacement. This procedure is repeated for each refractive index value to model the refractive index dependence of the trap stiffness (Fig.\;\ref{trace_index_wide}). The non-linear dependence originates from the interference structure of Mie scattering, which is described in detail elsewhere \cite{mazolli2003}. In both theory and experiment, transverse displacements along the direction of polarization ($x$) were considered. The trap stiffness $\alpha_x$ is given as the trapping efficiency $Q$; multiplication with $n_{\mathrm{med}}P/c$, where $P$ is the laser power and $c$ the speed of light, gives $\alpha_x$ in newtons/m.

\begin{figure}[t!]
   \includegraphics[clip,width=0.8\hsize]{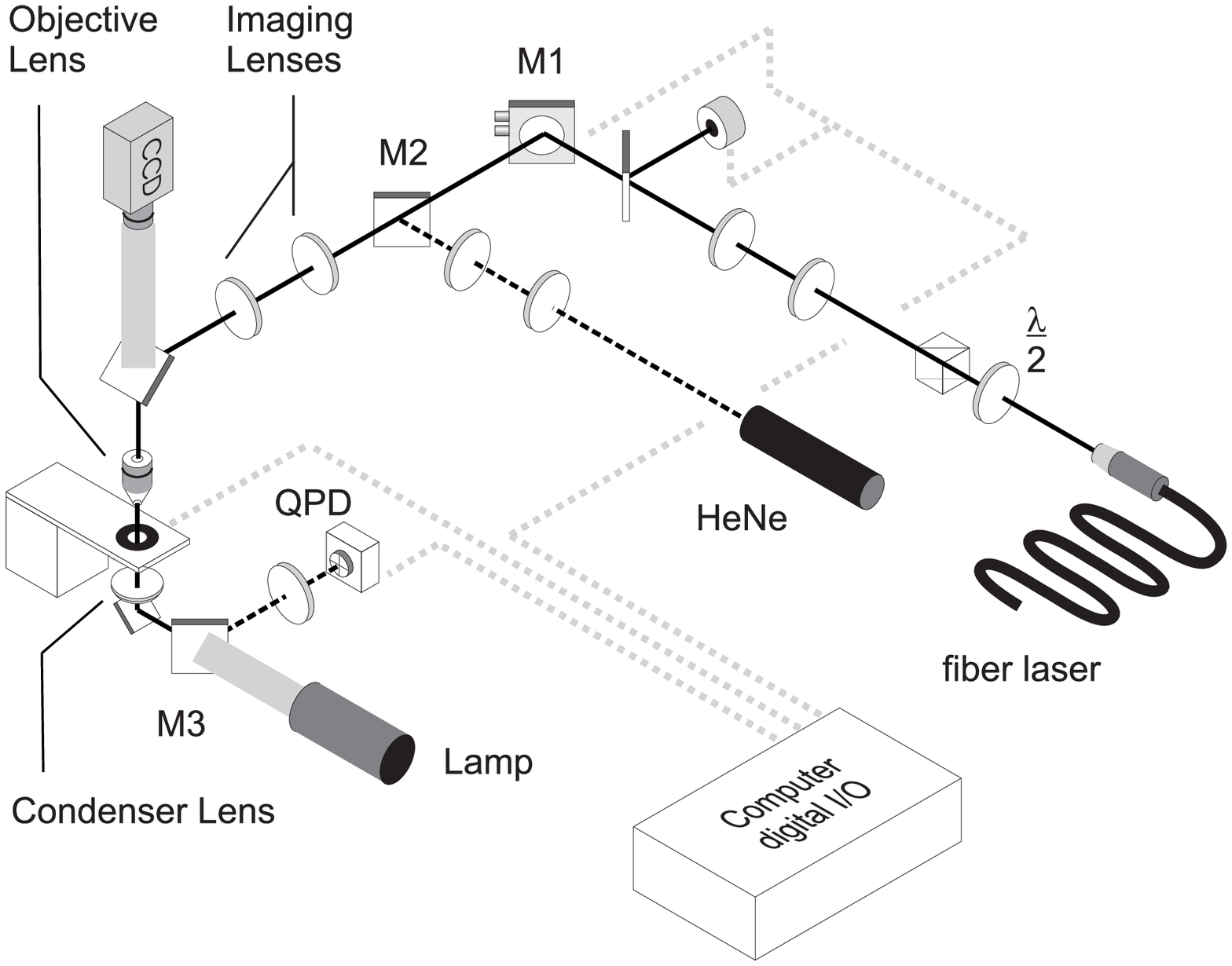}\\
   \includegraphics[clip,width=0.8\hsize]{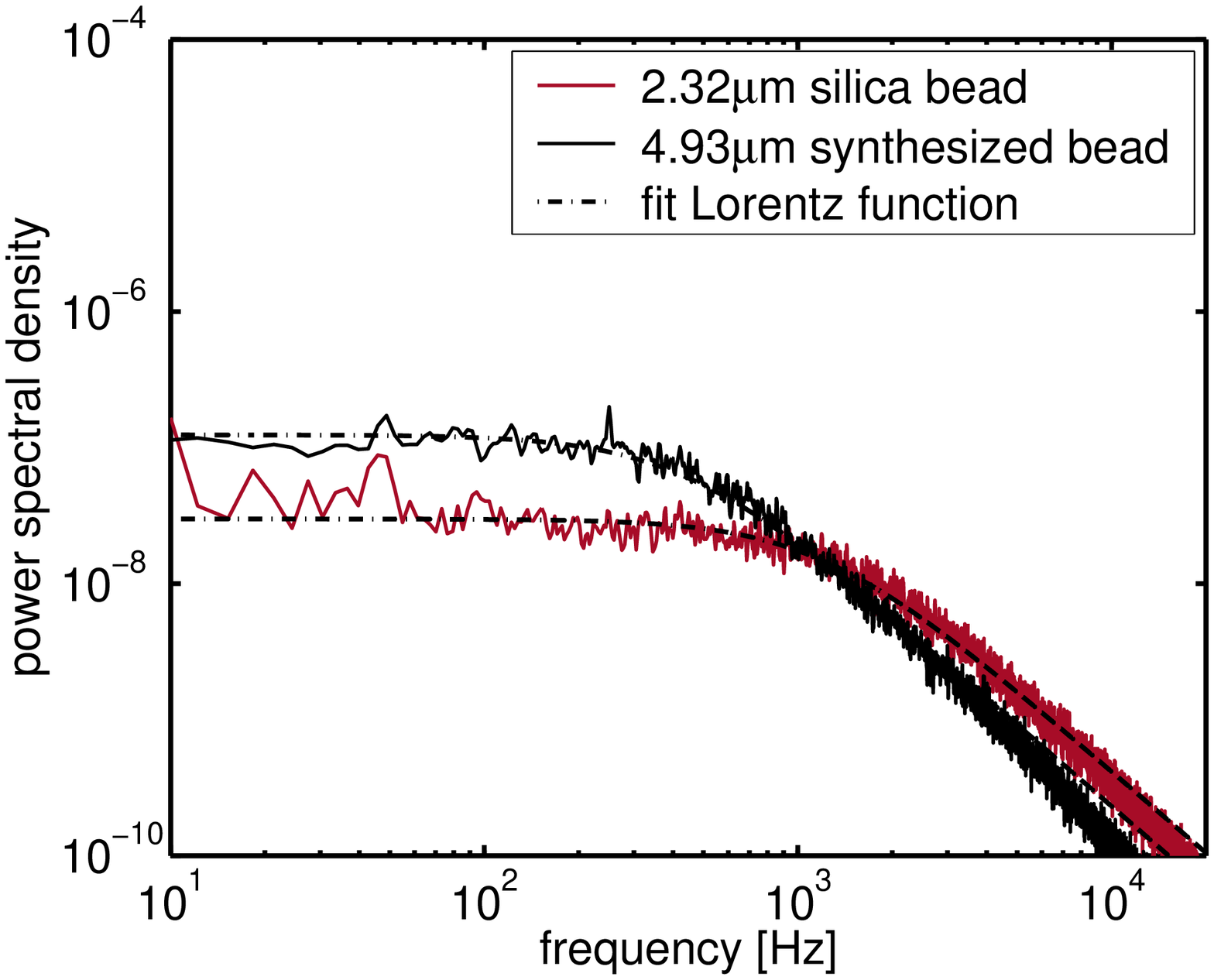}
   \caption{Upper: Setup used for trap stiffness measurements. Lower: Measured power spectral density of the thermal position fluctuations of particles in the optical trap. The roll-off frequencies are proportional to the trap stiffness.}
   \label{powerspectra_rob_si}
\end{figure}

The experimental setup to measure the transverse trap stiffness $\alpha_x$ is composed of a 1070\,nm fiber laser for trapping of the particle and a 633\,nm HeNe laser used in combination with a quadrant photo detector for particle position detection \cite{knoner2005}. The particle undergoes thermal motion in the trap. The power spectral density of its position fluctuations $|x(f)|^2\sim (f_0^2 + f^2)^{-1}$ has a typical roll-off frequency $f_0=\alpha_x /2\pi\beta$, which is obtained from a curve fit and used to calculate $\alpha_x$ (Fig.\;\ref{powerspectra_rob_si}). $\beta$ is the drag coefficient. For the experiments, silica (SI) particles with a diameter of $d=2.32$\,$\mu$m, poly(methyl methacrylate) (PMMA) particles with $d=1.68$\,$\mu$m (both Bangs Laboratories Inc., Fishers, IN) and polystyrene (PS) particles with $d=2.09$\,$\mu$m (Polysciences Inc. Warrington, PA) were obtained, washed and suspended in deionized water. Synthesized sulfurated silica particles (synSI) with a size range of $d=4$ to 6\,$\mu$m were kindly provided by the Department of Chemistry (The University of Queensland, Australia). Particle suspensions were sealed in a sample chamber of 50\,$\mu$m thickness. Stiffness measurements were performed in the center of the chamber, far away from interfering walls.

To reduce the experimental error, the trap stiffness was measured several times for each individual particle and for 2--3 particles for each type of particle. For PS, PMMA and SI, bead diameters were given by the manufacturer. For the synthesized polydisperse silica particles, particle diameters were determined by video microscopy. The optical image of microscopic particles is broadened due to diffraction, but the center to center distance of two touching particles is not affected. A more precise determination of the radius $r_1$ is achieved by using the optical trap to bring two particles of similar size into contact and measuring the center to center distance $d_c$ and their relative apparent diameter $d_{a2}/d_{a1}$:
$r_1=d_c/(1+d_{a2}/d_{a1})$.

\begin{figure}[t!]
   \includegraphics[clip,width=0.8\hsize]{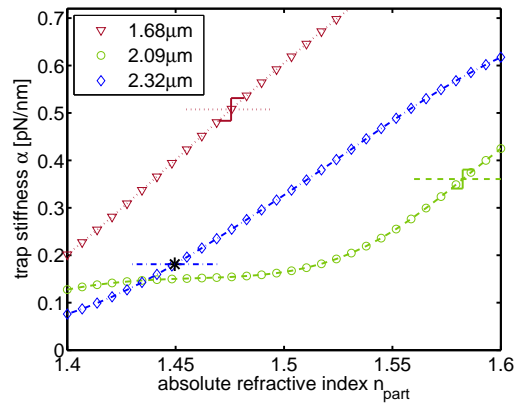}
   \caption{The refractive index of particles of the given size is determined from the intersection of the measured trap stiffness (horizontal lines) with the modeled trap stiffness (curves). The stiffness measured for SI (2.32\,$\mu$m) was used for calibration ({\bf *}).}
   \label{result_pmma_ps}
\end{figure}

For each particle size, the trap stiffness was calculated as a function of the particle refractive index using $n_{\mathrm{med}}=1.33$ \cite{irvine1968} (Fig.\;\ref{result_pmma_ps}). These curves do not have any free parameters; all parameters can in principle be experimentally accessed. Yet it is very difficult to measure the laser power at the focal spot. We precisely determine that parameter by a one-time calibration measurement with a well characterized particle (SI microsphere).

The refractive indices of the PS and PMMA particles were determined from the measured $\alpha_x$ (table \ref{table_index}) and the theoretical curves (intersections in Fig.\;\ref{result_pmma_ps}). The error was calculated from the variance of the different trap stiffness measurements and the resulting uncertainty in determining the refractive index. A 5\% error in the trap stiffness results only in a 0.2-0.5\% error in the refractive index because of the steep slope of the curves in Fig.\;\ref{result_pmma_ps} (note the $x$-axis range).
The error in the theoretical curves due to error in size and in calibration was also taken into account. The total error in the refractive index determination for these particles was less than 1\,\%.

\begin{table}[b!]
\begin{center}
  \begin{tabular}{cccc} \hline
    material & stiffness [pN/nm] & $n_{\mathrm{part}}$ measured & deviation [\%] \\ \hline\hline
    SI  &  $0.181\pm0.008$  &  $1.450$  & 0 (reference) \\ \hline
    PS  &  $0.361\pm0.019$  &  $1.582\pm0.011$  &  0.56 \\ \hline
    PMMA  & $0.507\pm0.024$  &  $1.476\pm0.012$  &  0.35 \\ \hline
  \end{tabular}
\end{center}
  \caption{Results of the refractive index determination and deviation from manufacturer/literature values.}
  \label{table_index}
\end{table}

For comparison, refractive index values for the different materials were obtained from the manufacturers and extrapolated to the wavelength of 1070\,nm by using Cauchy and Sellmeier dispersion relations \cite{ishigure1996,nikolov2000,ma2003,mellesgriot2002_2}. The measured values agree very well with the nominal values, the deviation being less than 0.6\,\% (table \ref{table_index}). The results show that high accuracy refractive index measurements can be made when probing single particles. The precision of the measurement could be even further enhanced by combining calibration particle and particle of interest in the same sample chamber and by averaging over a larger number of particles.

\begin{figure}[t!]
   \includegraphics[clip,width=0.8\hsize]{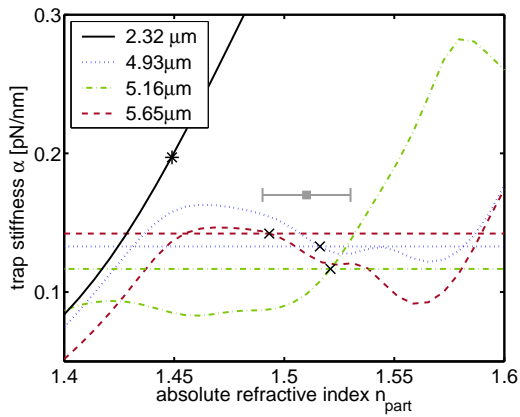}
   \caption{Experimental results of trap stiffness measurements for different sized synSI particles (horizontal lines) and corresponding theoretical curves. Intersections used for the refractive index evaluation are marked ({\sf x}). SI was again used for calibration ({\bf*}).}
   \label{result_si_rob}
\end{figure}

The refractive index of three synthesized silica particles was measured in a similar way. The larger particle sizes ($d_1=4.93\pm0.05$\,$\mu$m, $d_2=5.16\pm0.05$\,$\mu$m, $d_3=5.65\pm0.05$\,$\mu$m) cause the dependence of $\alpha_x$ on the refractive index to be non-unique for $d_1$ and $d_3$ (Fig.\;\ref{result_si_rob}). The shift of the steep linear region with size (Fig.\;\ref{trace_index_wide}) allows selection of a particle size for which the curve does increase monotonically in the region of interest (here $d_2=5.16\,\mu$m) and derivation of the particle index.

Multiple solutions for the measured $\alpha_x$ do not pose a problem. First, they occur only for large particle sizes. Second, a particle with a unique solution can usually be found. If that is not the case, then, third, one can combine the results for a number of particles with multiple solutions. The one common solution they all have will be the valid one. That is similar to a system of equations with multiple solutions, where the correct solution has to fulfill all the equations. Still, particles with one unique solution are always favorable because their curves have a steeper slope resulting in a smaller error.

A mean refractive index of $n=1.51\pm0.02$ is determined from the marked ({\sf x}) solutions (Fig.\;\ref{result_si_rob}). Extrapolation along the SI dispersion curve gives a value of $n=1.519\pm0.02$ at 589\,nm. That compares very well with the value of $n=1.523$ at 589\,nm for the same particles obtained from transmission spectroscopy using the method in ref. \cite{alupoaei2004}. The good agreement shows that the method described here is also suitable for determining the refractive index of larger polydisperse particle suspensions, and that polydispersity is even an advantage since it allows selection of a particles size with a favorable refractive index--stiffness relationship.

Other methods for single particle refractive index determination are largely based on the angular scatter pattern. Measurement of that pattern by photographic recording \cite{ashkin1980}, semicircular rotating stages \cite{doornbos1996} or an array of optical fibres \cite{ulanowski2002} prevent the method from being easily integrated into a standard microscope and to be compatible with microfluidic devices or lab-on-a-chip applications. The method described here does not suffer from any of these shortcomings as it requires only two dichroic mirrors to be added to a standard microscope. The technique is also compatible with microfluidic devices and lab-on-a-chip applications and thus allows {\it in situ} or {\it in statu nascendi} index measurements.

Other system parameters will be accessed in future applications of the method. With a well characterized probe particle, the refractive index of a fluid can be measured. For a system with known refractive indices, the stiffness dependence on size can be calculated and accurate size measurements obtained. Our recent advances in modeling of optical trapping of non-isotropic and non-spherical particles will allow us to extend the technique to more complex systems in the future.

In conclusion, we have developed and demonstrated a novel technique for the measurement of the refractive index of single microscopic particles. The technique has potential for application to a range of problems and is compatible with standard microscopy and microfluidic devices. The applicability of rigorous modeling of optical tweezers in the strong interference regime (1-5\,$\mu$m) for a wide parameter range is demonstrated.

\begin{acknowledgments}
The contributions of Katrina Seet and Dr. Robert Vogel, in particular particle synthesis and transmission spectroscopy, are greatly acknowledged.
\end{acknowledgments}


\end{document}